\documentclass[12pt]{article}
\usepackage[utf8]{inputenc}
\usepackage{graphicx}
\usepackage[margin=1in]{geometry}
\usepackage[dvipsnames]{xcolor}
\usepackage{authblk}
\usepackage{amsmath, amssymb}
\usepackage{multirow}
\usepackage{wrapfig} %tighter figures
\usepackage{placeins} % has \FloatBarrier command that will keep figures from winding up after refs
\usepackage{url}

\title{\textbf{Geometries and fabrication methods for 3D printing ion traps}}

\author[1]{A. Quinn}
\author[2]{M. Brown}
\author[2]{T.J. Gardner}
\author[1]{D.T.C. Allcock}
\date{\today}

\affil[1]{Department of Physics, University of Oregon, Eugene, Oregon 97403, USA}
\affil[2]{Knight Campus, University of Oregon, Eugene, Oregon 97403, USA}

\begin{document}

\maketitle

\section*{Abstract} The majority of microfabricated ion traps in use for quantum information processing are of the 2D `surface-electrode' type or of the 3D `wafer' type. Surface-electrode traps greatly simplify fabrication and hold the promise of allowing trapped-ion quantum computers to scale via standard semiconductor industry fabrication techniques.  However, their geometry constrains them to having much lower trapping efficiency, depth, and harmonicity compared to 3D geometries.  Conversely 3D geometries offer superior trap performance but fabrication is more complex, limiting potential to scale. We describe new `trench' geometries that exist in the design space between these two paradigms.  They still allow for a simple, planar electrode layer but with much more favourable trapping properties.  We propose such traps could be 3D-printed over a 2D wafer with microfabricated components already integrated into it, thus retaining all the integration techniques and scaling advantages of surface-electrode traps. As a proof of principle we use 2-photon direct
laser writing lithography to print the required electrode structures with the proposed geometry.

%This represents the first application of additive 3D printing technology to ion trap fabrication.

%Large design space in between wafers and surface that are amenable to monolithic microfabrication.  Wafer hard to realize - already somewhat of an approx. Symmetries.

\section{Introduction} 

%The trapped-ion platform has been at the forefront of quantum information processing since its first demonstration in the mid-90s.  Its continued relevance though will depend on whether current barriers to scaling can be surmounted.  Whilst all platforms have many such barriers, they are considered particularly acute in trapped ions because of the requirement for both optical and electronic classical control.

\begin{figure}[h]
\includegraphics[width=1\textwidth]{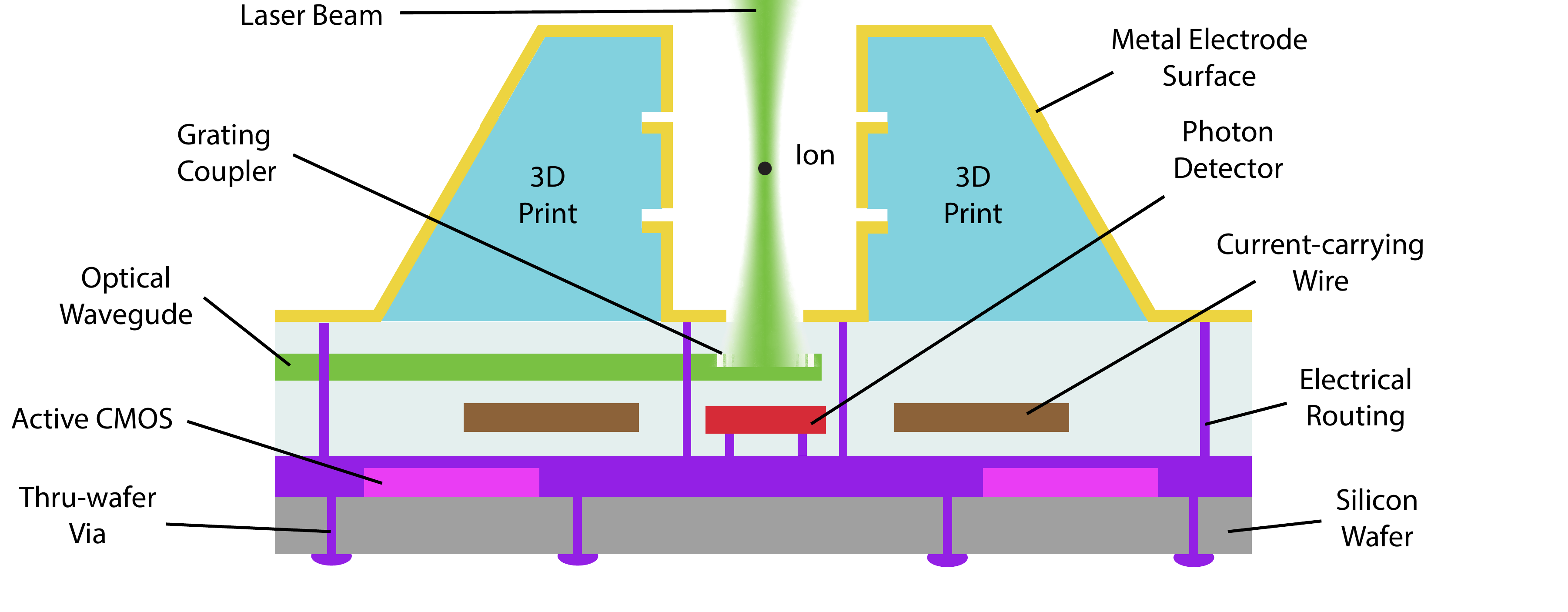}
\centering
\caption{\label{wafer}Cross-section of a trap fabricated using our proposed method of 3D printing an electrodes onto a wafer containing electrode routing and some of the potential integrated ion control elements that have been demonstrated.}
\end{figure}

The surface-electrode ion trap (SET)~\cite{Chi05} was a major step forward for realizing the trapped ion quantum computer when it was first demonstrated in 2006~\cite{Sei06}.  The SET allows for the trap electrodes to be fabricated as a single layer on a wafer.  This has allowed for a diverse range of traps to be easily fabricated with rapid turnaround in relatively simple university cleanrooms as well as with commercial MEMS and CMOS processes~\cite{Blain_2021}. Underneath this top electrode layer can be many other functional layers containing all the required classical control elements, as shown schematically in Fig.~\ref{wafer}.  These include current-carrying wires for magnetic field-driven gates~\cite{Allcock13,Warring13}, photonic integrated circuits for laser delivery~\cite{Mehta2016, Ivory21, Mehta20}, superconducting nanowires~\cite{Slichter17, Todaro_2021} or avalanche photodiodes~\cite{Setzer_2021, Reens_2022} for photon detection, active CMOS electronics such as DACs for trap voltage generation~\cite{Stuart19}, and routing layers and thru-wafer vias~\cite{Guise15} to external connections.  Whilst the complete set of control elements required for QIP has yet to be integrated into a single  device, they have all been demonstrated separately and efforts to combine them are underway.

% Fig. 4 inset is just for the patent people

3D `wafer' traps on the other hand typically require thru-wafer machining~\cite{Wilpers12}, and usually alignment and stacking of multiple wafers, to achieve the desired geometries.  These processes generally preclude the CMOS-like monolithic approach to integrating functional elements, and instead relies on further serial assembly of bespoke and heterogeneous components that requires much new process development~\cite{Ernzer, AQTION, Translume, Day21}.  The recently developed approach of combining wafer stacking and SETs~\cite{Auchter22} potentially offers a solution to some of these issues.

%In large part because of these drawbacks, many research groups still use 3D traps extensively and work to develop ways to integrate classical control into serially-assembled 3D-traps~\cite{Ernzer, AQTION, Translume}.  This is despite their limited scaling potential compared to monolithic traps fabricated on wafers using industry-standard processes (specifically CMOS and CMOS-compatible processes). 

Despite these fabrication disadvantages, wafer traps are still under active development because SETs have several major drawbacks~\cite{Blakestad09}, specifically:
\begin{itemize}
\item  Trapping potentials are less harmonic.  This causes any slight displacement of the trapping location due to stray charges or control voltage inaccuracies to shift the trap frequencies.  Without recalibration this will lead to errors in multi-qubit gates.  The increased cross-Kerr nonlinearity can also lead to gate errors due to spectator modes~\cite{Roos_2008, Nie_2009}.
\item  Trap depths are much lower~\cite{Chi05}, typically close to room temperature.  This makes loading ions less efficient and ion losses due to control field errors, stray charges, and background gas collisions are much more likely. Room temperature operation of more than a couple of ions in lighter ion species is effectively precluded due to short ion lifetimes.  
\item Junctions between ion traps with favourable trapping strength and depth, and small pseudopotential barriers are harder to achieve than with a 3D geometry~\cite{Wesenberg2009}.
\item  As the ions reside in an open volume above a surface, they are poorly shielded from electric charges or crosstalk (photons, microwave control fields, or electric trapping fields) from other trapping zones on the SET. 
\end{itemize}

In this paper we propose mitigating these issues by adding a 3D-printed dielectric structure onto the trap wafer before depositing the final metal electrode surface.  This allows us to retain the fabrication advantages of the SET whilst also retaining the advantageous trapping geometry of a 3D trap.  This requires finding trap geometries that meet these requirements.  In section~\ref{quantifying_trap_performance} we lay out how to quantify the performance of a given geometry.  In section~\ref{trench_trap_geometries} we propose several feasible geometries and simulate them.  Then finally, in section~\ref{3d_printing} we identify a suitable 3D printing process and print demonstration devices.

\section{\label{quantifying_trap_performance} Quantifying Trap Performance}

As alternatives to existing microfabricated trap designs, we explore several possible `trench' geometries, discussed in greater detail in the following section.  To assess the viability of these designs, we simulate them with COMSOL Multiphysics using the boundary element method (BEM). The Electrode software package~\cite{Electrode} is used for further post-processing of the electric fields extracted from the simulations.

\begin{figure}[h]
\includegraphics[width=0.8\textwidth]{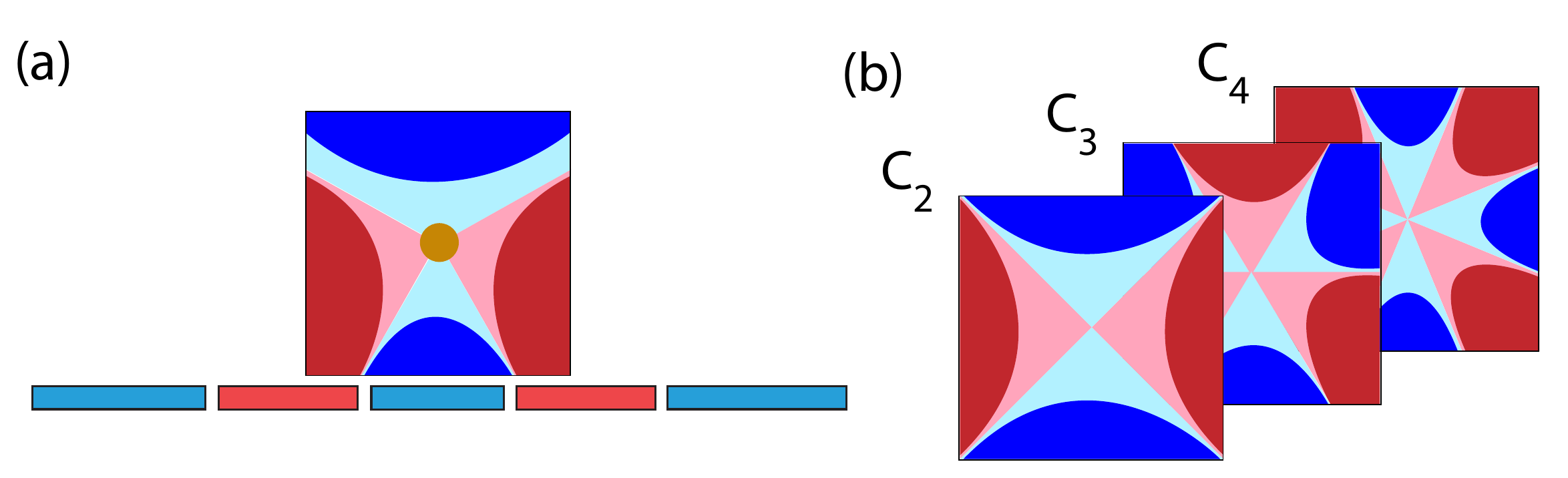}
\centering
\caption{\label{multipole}(a) A schematic cross-section of the potential generated by a SET, with the equilibrium position of the ion marked with a dot. (b) schematic cross-sections of the first three multipoles the potential in (a) can be expanded into, labelled with their respective expansion coefficients.}
\end{figure}

In each trap simulation we set RF electrodes to unit voltage and DC electrodes to ground and calculate the electric field generated in the vicinity of the ion.  From these fields, we can calculate the pseudopotential of the trap, which is the effective potential seen by the ion~\cite{DEHMELT1968}.  By finding where this pseudopotential is at a minimum in the plane perpendicular to the trap axis, we locate the equilibrium position of the ion in two dimensions.  By also finding the saddle point of the pseudopotential (ie, the `escape point') and comparing the pseudopotential at this point to the pseudopotential at the center of the trap, we have a rough proxy for the depth of the trap (though in reality the situation is more complex due to the dynamical nature of a Paul trap).  At the ion's position, we fit the surrounding electric potential $V=V(r, \theta)$ to a cylindrical harmonic expansion

\begin{equation}
    V(r,\theta) = V_0\sum_{n=2}^\infty C_n \left(\frac{r}{r_0}\right)^n \cos(n\theta + \phi_n) + V_{off}
\end{equation}

\noindent where $r$ is distance from the trap center, $r_0$ is the minimum ion-electrode separation, $V_{off}$ is the potential at the ion, $\phi_n$ is the orientation of the $n$th-order multipole, and $C_n$ is the expansion coefficient associated with an $n$th-order multipole.  From these expansion coefficients, we define what we call the `quadrupole,' `hexapole,' and `octopole' strengths generated by our traps,
\begin{align*}
    \text{quadrupole}, C_2, \\
    \text{hexapole}, C_3' =& \frac{C_3}{C_2}, \\
    \text{octopole}, C_4' =& \frac{C_4}{C_2},
\end{align*}

\noindent which we illustrate schematically in Fig.~\ref{multipole}.

The quadrupole term gives the geometric efficiency of the trap.  Greater efficiency is desirable as it reduces the required RF voltage, and thus power dissipation as well as the possibility of voltage breakdown on the chip.  The hexapole and octupole terms give the leading order anharmonicities in the potential and are normalized to the quadrupole term as it is most useful to compare traps at the same ion secular frequency, not the same RF voltage.  Anharmonicities  increase cross-Kerr nonlinearity and introduce displacement-dependent frequency shifts in the ion's motion, increasing sensitivity of gate fidelity to the temperature of non-gate modes and the presence of stray electric fields.  Therefore, we seek designs which minimize higher-order multipole terms around the trap center.  %with small hexapole and octopole terms.  We consider only these leading-order anharmonicities because, for ion displacements $r$ much smaller than ion-electrode separation $r_0$, the contribution of the $n$th term is of $r/r_0$ weaker than that of the $(n-1)$th term, for $C_n \sim C_{n-1}$.  %For a more complete discussion of the effects of anharmonic terms on trap frequencies, see the Appendix.%This means that, for example, if a potential has comparable hexapole, octopole, and decapole terms $C_3 \sim C_4 \sim C_5$, then for a relatively large displacement $r/r_0 \sim 1/10$, the octopole will have an effect $\sim 1/10$ as large as the hexapole's, while the decapole will have an effect $\sim 1/100$ as large.  

In addition to trap depth and the multipole components of the trapping potential, we also calculate the open numerical aperture (NA) available for fluorescence collection and for laser beam access, as illustrated in Fig.~\ref{designs}.  A high NA (typically $\sim0.4$) is important for fast state readout on the ion.  For the novel designs discussed in Sec.~\ref{trench_trap_geometries}, we give an NA from above and from the substrate, as the ion detectors could be located in the substrate~\cite{Slichter17, Todaro_2021, Setzer_2021}, or placed above the trap.  For SETs the NA is trivially one, while our wafer traps, by their symmetry, have the same NA from both sides.  In all cases we consider a circular detector, but a modest improvement could be achieved with a rectangular detector.  These NA also gives us an idea of the range of angles available for laser beams perpendicular to the trap axis such that they can exit the trap without striking an electrode.

%\begin{figure}[h]
%\includegraphics[width=0.8\textwidth]{simulation.pdf}
%\centering
%\caption{\label{simulation}Sample output of a BEM simulation performed in COMSOL of the stacked trench design (see Sec.~\ref{trench_trap_geometries}), showing surface charge density, cross section of electric potential in free space, and model mesh.  Ion position is marked with a yellow dot.}
%\end{figure}

We scale all of our traps such that the closest electrode to the ion is 75\,$\mu$m away.  Because anomalous heating has a strong $\sim d^4$ dependence~\cite{King_1998, Turchette_2000}, the closest electrode will tend to dominate the heating rate.  For calculating trap depths we assume a $^{40}$Ca$^+$ ion trapped at a secular frequency of 4\,MHz by a 40\,MHz RF trap drive with no DC potentials applied.

\section{\label{trench_trap_geometries} Trench Trap Geometries}

\begin{figure}[h]
\includegraphics[width=0.9\textwidth]{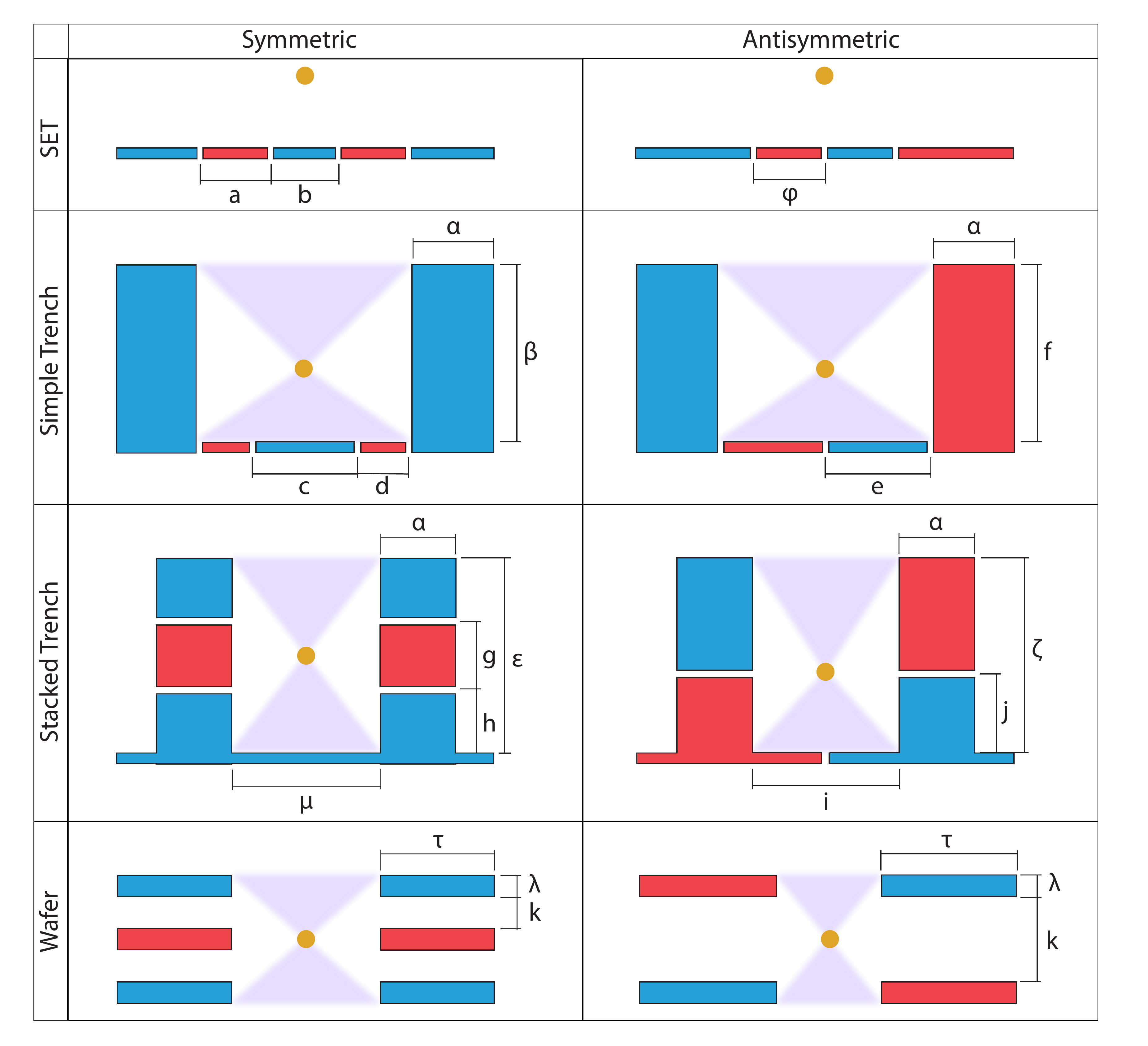}
\centering
\caption{\label{designs}Cross-sectional views of of the trap designs simulated.  The ions' location is shown as an orange dot and the trap axis is out of the page.  DC electrodes are blue, RF electrodes are red.  Numerical apertures are shown in lavender.}
\end{figure}

3D printing offers great geometric versatility in trap design, since it can be used to produce arbitrary shapes.  In principle, we could 3D print electrodes with hyperbolic cross-sections.  Such electrodes would generate pure quadrupole potentials, with lower anharmonicities (as discussed in Sec.~\ref{quantifying_trap_performance}) than other geometries.  In practice however, our choices of geometry are limited by other fabrication and operation constraints, namely whether the geometry allows optical access to the ion for laser beam delivery and fluorescence detection and how easily the 3D-printed dielectric can be metalized (a problem discussed further in Sec.~\ref{3d_printing}).  These constraints naturally suggest what we refer to as a `trench' geometry, illustrated in Fig.~\ref{wafer}.  Trench geometries confine ions between two walls 3D printed on the surface of a wafer.  We explore several such `trench' geometries for microfabricated ion traps and compare their performance to SETs and wafer traps of similar dimensions.  Cross-sectional views of the geometries simulated are shown in Fig.~\ref{designs}.

The trap geometries we considered can be classified as either symmetric or anti-symmetric. Both sets of trap geometries are mirrored around a vertical plane along the trap axis, but in the anti-symmetric traps the role of RF and DC electrode is switched in the mirror half of the trap.  Obviously many asymmetric variations on these traps are possible, usually with properties somewhere between the symmetric or anti-symmetric ones.  We do not consider them here though as symmetries are generally desirable, and pragmatically we need to keep the parameter space of different traps to simulate tractable. 

As shown in Fig.~\ref{designs}, all simulated traps aside from the SET are characterized by $\ge 3$ geometric parameters.  We select a subset of parameters (marked with Latin letters) to sweep while fixing the other parameters (marked with Greek letters).  For our trench traps, we also investigate how performance changes with the more relevant of the fixed parameters.  We do so by making multiple sweeps over the continuous parameters for different values of a fixed parameter.  %We sweep over only one parameter for designs where keeping ion-electrode separation at 75\,$\mu$m is trivial, whether this is because the ion position is set by simple symmetries (as with the wafer traps) or because the ion position can be calculated analytically (as with the SET).  For our trench traps (aside from the symmetric stacked trench), the situation is more complicated, since ion position is not immediately clear from the trench geometries and since we do not have an analytic expressions for trap positions in terms of the geometric parameters.  Therefore, to find geometries that will satisfy our constraint on ion-electrode separation, we need to carry out full sweeps over a pair of parameters (e.g. $c$ and $d$ for the symmetric simple trench) and find parameter combinations for which the constraint is satisfied, which amounts to finding an isoline in height over the 2D parameter space.  We then characterize the trench trap along this isoline.  

%As stated in the previous section, all of our traps are scaled such that the minimum ion-electrode separation is 75\,$\mu$m.  This means that, for any sweep over two parameters, we are only interested in the pairs of parameter values for which the trap meets this condition.  Finding these parameter values is trivial for some trap designs, i.e. the wafer traps, making sweeps over a 2D parameter space unnecessary.  The wafer traps can be characterized with a sweep over a single parameter.

\subsection{Simple Trench}

The first trap geometry we consider is the `simple trench'.  In this design, the two outermost electrodes of the SET are extended vertically to above the height of the ion.  The symmetric simple trench is primarily characterized by four dimensions, DC and RF electrode widths $c$ and $d$, trench wall height $\beta$, and trench wall thickness $\alpha$.  This last dimension does not strongly affect the potential seen by the ion, since the ion is shielded from all but the inner surfaces of the trench.  Because of this, a fixed value $\alpha = 100$\,$\mu$m, comparable to values used in our test fabrications (see section~\ref{3d_printing}), is used in all trench trap simulations.

\begin{figure}[h]
\includegraphics[width=1.0\textwidth]{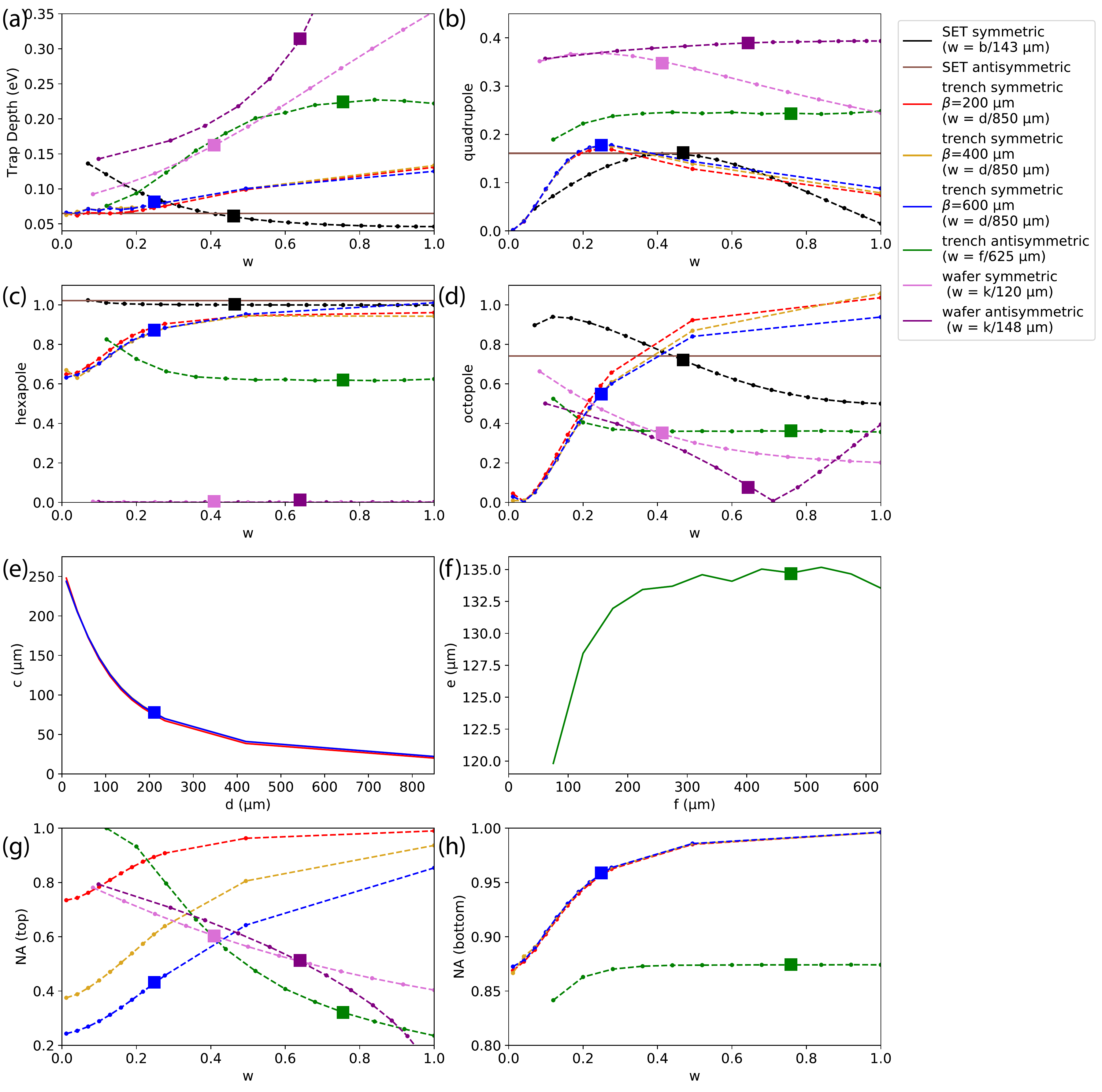}
\centering
\caption{\label{simple_trench}Simulated parameters of SET traps vs the symmetric and anti-symmetric `simple trench' geometry. (a) Trap depth at a fixed radial frequency. (b) The quadrupole component ($C_2$) of the trapping potential at the ion.  (c) The hexapole component ($C_3'$) of the trapping potential.  (d) The octopole component ($C_4'$) of the trapping potential.  (e) Electrode dimensions at a constant ion-electrode separation for the symmetric simple trench trap.  (f) Electrode dimensions at a constant ion-electrode separation for the anti-symmetric simple trench trap. (g,h) NA above and below the ion.  Parameters are plotted against a generic geometric variable $w$, which is related to an electrode dimension of each trap as shown in the legend.  Simulation results are shown with markers.  The representative SET, wafer, and simple trench traps whose characteristics are summarized in Table \ref{tab:trap-summary} are highlighted with larger markers.  Simulation points are connected with dashed lines only as a guide to eye.}
\end{figure}

We simulate a set of symmetric simple trench traps at a fixed ion height, allowing $c$ and $d$ to vary for discrete values of $\beta$.  Unlike the symmetric design, the anti-symmetric simple trench can be characterized with just two parameters, plane electrode width $e$ and trench wall height $f$.  (We find that ion height in the anti-symmetric trench depends mainly on $e$, as shown in Fig.~\ref{simple_trench}(f).  In fact, past $f\approx200$\,$\mu$m, ion height shows no dependency on $f$ noticeable beyond numerical noise.)  The full simulation results for our symmetric and anti-symmetric simple trench traps are shown in Fig.~\ref{simple_trench}.

%\begin{figure}[h]
%\includegraphics[width=1.0\textwidth]{data_trench_sym_marked.pdf}
%\centering
%\caption{\label{sym_simple_trench}Simulated parameters of SET traps vs the symmetric `simple trench' geometry. Parameters are plotted against a generic geometric variable $w$, which is related to an electrode dimension of each trap as shown in the legend.  The representative SET and symmetric simple trench trap whose characteristics are summarized in Table \ref{tab:trap-summary} are marked with squares.}
%\end{figure}

%\begin{figure}[h]
%\includegraphics[width=1.0\textwidth]{data_trench_anti_marked.pdf}
%\centering
%\caption{\label{anti_simple_trench}Simulated parameters of SET traps and wafer traps vs the anti-symmetric `simple trench' geometry.  Parameters are plotted against a generic geometric variable $w$, which is related to an electrode dimension of each trap as shown in the legend.  The representative SET and anti-symmetric simple trench trap whose characteristics are summarized in Table \ref{tab:trap-summary} are marked with squares.}
%\end{figure}

\subsection{Stacked Trench}

The simple trench geometry discussed above is closer to the ideal quadrupole geometry than a surface trap, but the flexibility of this geometry is limited by the fact that each wall can only contain one electrode.  As an alternative, we consider what we call the `stacked trench' geometry.  By allowing multiple electrodes to make up the trench wall, we can produce arrangements that more closely match the boundary conditions of a quadrupole, though the fact that the trench must remain open on the top prevents any trench geometry from being fully rotationally symmetrical. 

The symmetric stacked trench trap can be characterized by four parameters: trench height $\epsilon$, trench width $\mu$, RF electrode height $g$, and the height of the RF electrode off of the ground plane $h$.  The anti-symmetric stacked trench trap can be characterized by three parameters: trench height $\xi$, bottom electrode height $j$, and trench width $i$.

These traps have two regimes for which ion-electrode separation is  75\,$\mu$m.  In the first regime, the ion is 75\,$\mu$m from the ground plane and the trench walls are further away, while in the second regime, the ion is 75\,$\mu$m  from the trench walls.  We consider both regimes in the antisymmetric case, but only the second in the symmetric case, since performance in the other regime was notably worse.

\begin{figure}[h]
\includegraphics[width=1.0\textwidth]{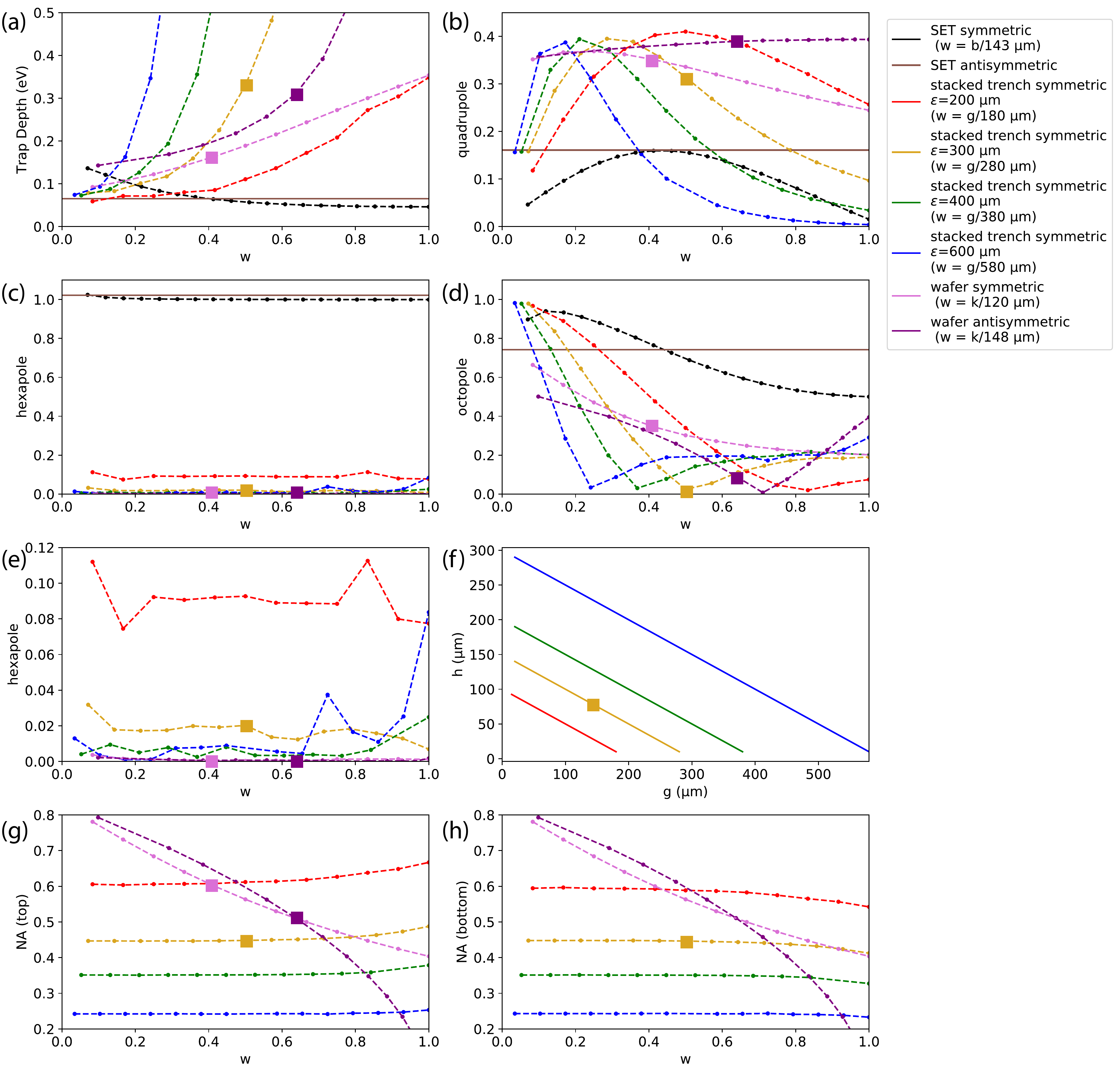}
\centering
\caption{\label{sym_stacked_trench}Simulated parameters of SET traps and wafer traps vs the symmetric `stacked trench' geometry. (a) Trap depth at a fixed radial frequency. (b) The quadrupole component ($C_2$) of the trapping potential at the ion.  (c) The hexapole component ($C_3'$) of the trapping potential.  (d) The octopole component ($C_4'$) of the trapping potential.  (e) The hexapole component of the trapping potential, excluding the SET data to make the other series more clear.  (f) Electrode dimensions at a constant ion-electrode separation for the symmetric stacked trench trap. (g,h) NA above and below the ion.   Parameters are plotted against a generic geometric variable $w$, which is related to an electrode dimension of each trap as shown in the legend.  Simulation results are shown with markers.  The representative wafer and symmetric stacked trench traps whose characteristics are summarized in Table \ref{tab:trap-summary} are highlighted with larger markers.  Simulation points are connected with dashed lines only as a guide to eye.}
\end{figure}

%\begin{figure}[h]
%\includegraphics[width=0.9\textwidth]{data_stacked_sym_d2.pdf}
%\centering
%\caption{\label{sym_stacked_trench_d2}Simulated parameters of a SET trap vs the `simple trench' geometry.}
%\end{figure}

\begin{figure}[h]
\includegraphics[width=1.0\textwidth]{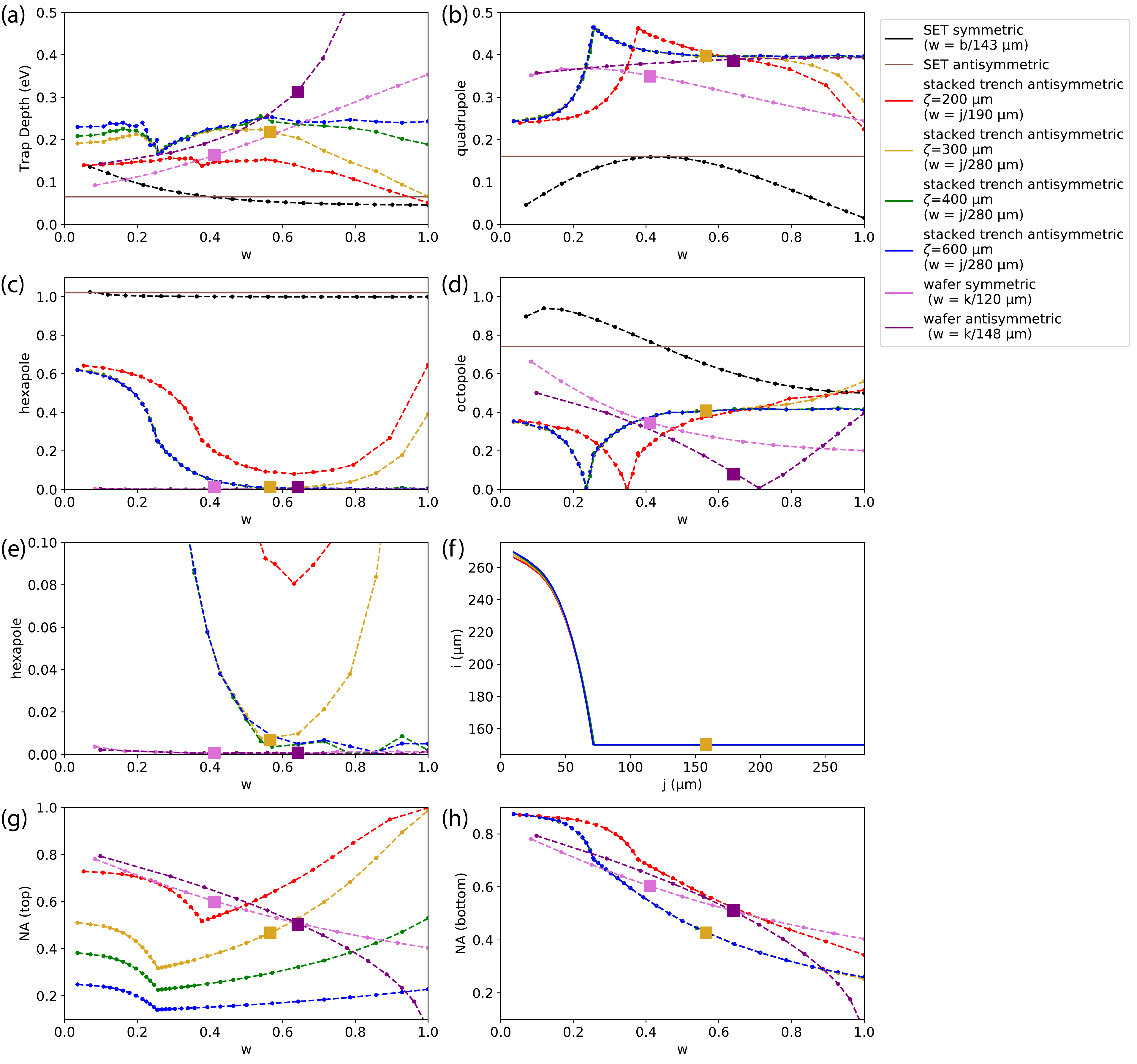}
\centering
\caption{\label{anti_stacked_trench}Simulated parameters of SET traps and wafer traps vs the anti-symmetric `stacked trench' geometry. (a) Trap depth at a fixed radial frequency. (b) The quadrupole component ($C_2$) of the trapping potential at the ion.  (c) The hexapole component ($C_3'$) of the trapping potential.  (d) The octopole component ($C_4'$) of the trapping potential.  (e) The hexapole component of the trapping potential, excluding the SET data to make the other series more clear.  (f) Electrode dimensions at a constant ion-electrode separation for the anti-symmetric stacked trench trap. (g,h) NA above and below the ion.   Parameters are plotted against a generic geometric variable $w$, which is related to an electrode dimension of each trap as shown in the legend.  Simulation results are shown with markers.  The representative wafer and anti-symmetric stacked trench traps whose characteristics are summarized in Table \ref{tab:trap-summary} are highlighted with larger markers.  Simulation points are connected with dashed lines only as a guide to eye.}
\end{figure}

%We find different characteristics from the simple trench by allowing multiple electrode layers to be embedded in the trench walls.  While the symmetric and anti-symmetric stacked trenches we consider have only three and two layers respectively, we can easily imagine a trench with more layers.  However, for harmonic trapping, we want to generate a quadrupole potential, and adding e.g. another RF electrode layer to the symmetric stacked trench would take the boundary conditions of the system further from those of a quadrupole and strengthen the undesired higher-order multipole components of the potential.  Therefore, we do not consider designs with additional electrode layers.

\subsection{Comparison Geometries - SET and Wafer Traps}

As a point of comparison, we simulate common varieties of microfabricated trap: the SET and the wafer trap.  We first simulate the symmetric and anti-symmetric SET (also known as 5-wire and 4-wire SETs respectively).  One nice feature of SETs is that under the assumption of an infinitely large trap with no gaps between electrodes, simple closed-form expressions for the potential are available~\cite{House2008}.  While we use these closed-form expressions to calculate the parameters shown in Figs.~\ref{simple_trench}-\ref{anti_stacked_trench}, we also simulate our SETs in COMSOL as a way of benchmarking our BEM simulations.  We compare the multipole expansions, trap heights, and trap depths of BEM-generated and analytic potentials for the symmetric SET.  For most of the free parameter range, agreement across these parameters is within 2\%.  In the range where $a \ll b$ (see Fig.~\ref{designs}), divergence approaches 10\% due to the fact that the 1\,$\mu$m electrode spacing present in all BEM simulations is no longer negligible~\cite{Schmied_2010}.  For a fixed ion height there is one free parameter in the symmetric SET and no free parameters in the anti-symmetric one.

To compare our proposed trench designs against the most prominent set of current 3D microfabricated designs, we also simulate symmetric and anti-symmetric wafer traps, illustrated in the bottom row of Fig.~\ref{designs}.  A wafer thickness of $\lambda=50$\,$\mu$m of our simulated traps is comparable to those of designs implemented in the lab~\cite{Jost2010,Kienzler2015}.  The depth $\tau=1$\,mm of the wafers is selected to be much larger than wafer thickness and spacing so that the exact depth is unimportant.   Therefore, both designs have only one free parameter, the spacing $k$ between wafers in a stack.  This parameter is swept over while holding the ion-electrode distance constant.   %Comparing our own results for the symmetric and anti-symmetric wafer trap against earlier simulations performed with similar designs, we found that our results were consistent~\cite{Blakestad10}

\subsection{Summary}

Having generated the data shown above, we want to compare the simple and stacked trench trap designs with established SET and wafer trap designs and assess their relative merits.  For easy comparison between these four sets of simulated geometries, we select a `representative' trap from each category.  These traps in some qualitative way optimize the balance between trap depth and quadrupole strength/purity.  This selection is somewhat subjective, and optimum parameters may vary by application, but they are unlikely to differ greatly from what we have chosen here.  A table of values for there representative traps are shown in Table~\ref{tab:trap-summary}, and the associated dimensions given in Table~\ref{tab:trap-dims-summary}.

\begin{table}
\centering
\begin{tabular}{|c|c c|c c|c c|c c|} 
\hline
& \multicolumn{2}{|c|}{SET} & \multicolumn{2}{|c|}{Simple} & \multicolumn{2}{|c|}{Stacked} & \multicolumn{2}{|c|}{Wafer}  \\
\hline
           & Sym & Anti & Sym & Anti & Sym & Anti & Sym & Anti  \\
\hline
\hline
Depth (eV)         & 0.06 & 0.07 & 0.08 & 0.23 & 0.33 & 0.22 & 0.16 & 0.39 \\
\hline
Quadrupole ($C_2$) & 0.17 & 0.17 & 0.18 & 0.24 & 0.31 & 0.40 & 0.35 & 0.39 \\
\hline
Hexapole ($C_3'$)  & 1.0 & 1.0 & 0.86 & 0.62 & 0.020 & 0.008 & 0.000 & 0.001 \\
\hline
Octopole ($C_4')$  & 0.75 & 0.75 & 0.55 & 0.36 & 0.024 & 0.407 & 0.344 & 0.007 \\
\hline
\end{tabular}
\newline
\caption{Summary of results for trap parameter calculations for the SET, simple trench trap (`Simple'), stacked trench trap (`Stacked'), and wafer trap. Values shown correspond to a set of geometric parameters selected to balance trap depth and quadrupole strength/purity.}
\label{tab:trap-summary}
\end{table}

% Table of dimensions
\begin{table}
\centering
\begin{tabular}{|c|c|l|} 
\hline
\multicolumn{2}{|c|}{Trap} & Dimensions ($\mu$m) \\
\hline
\hline
\multirow{2}{*}{SET} & Sym  & $a=161.2$, $b=59$ \\
                     & Anti & $\phi=75$\\
\hline
Simple & Sym  & $d=77.3$, $c=210$, $\beta=600$ \\
                      Trench  & Anti & $e=135.2$, $f=525$ \\
\hline
Stacked & Sym  & $g=140$, $h=80$, $\epsilon=300$ \\
                        Trench & Anti & $i=150$, $j=160$ \\
\hline
\multirow{2}{*}{Wafer} & Sym  & $k=50$ \\
                       & Anti & $k=53$ \\
\hline
\end{tabular}
\newline
\caption{The dimensions (in microns) of the traps whose parameters are summarized in Table \ref{tab:trap-summary}.}
\label{tab:trap-dims-summary}
\end{table}

From Table~\ref{tab:trap-summary}, we predict that all our trench geometries will outperform SETs.  For the symmetric simple trench trap, the improvement is practically negligible though. The antisymmetric simple trench is a significant improvement over the SET, offering $\sim$3$\times$ the depth, $\sim$40\% higher quadrupole strength, and $\sim$40\% lower hexapole strength.  Despite this it is still notably inferior to the wafer traps. The stacked trench traps on the other hand can have very comparable performance to wafer traps, with similar depth, quadrupole strength, and octopole terms.  The hexapole term for our representative trap is still larger than in wafer traps (where it can be zero by symmetry), but the factor $\sim50$ improvement over SETs will still lead to drastic improvement in trap performance.  Further improvement can be gained at the cost of lower NA optical access by making the walls higher.  As some point though manufacturing tolerances and curvature from stray fields will set a practical limit for either design.  %Since the stacked trench trap is both a notable improvement on existing 2D designs, and should also be relatively easy to produce, we select this design for fabrication tests.

\section{\label{3d_printing} 3D Printing}

Our chosen fabrication process for realizing these trench geometries is 2-photon direct laser write (DLW) 
3D printing. The electrode structure is printed directly onto a functional trap wafer using a dielectric material.  The printing process uses a focused laser to polymerize this photosensitive material to generate solid structures~\cite{Pearre2019}. This can then be metallized to create conductive electrode surfaces.  

Whilst not explored here, we note that the simple trench traps could be fabricated using UV-LIGA (Ultraviolet - Lithographie, Galvanoformung, Abformung). This process is already widely used to make electrodes for SETs~\cite{Arr13} and so would be straightforward to adopt.  UV-LIGA  is a fabrication process where metal is electroplated inside a photoresist mold patterned using UV lithography. As discussed above, the performance advantage of these designs over SETs is fairly modest compared to the stacked trench traps.  The stacked trench geometries could also be built using conventional MEMS techniques.  However, the 100s of $\mu$m vertical heights required are likely challenging to achieve with oxide film deposition.\\  %Therefore we propose using 3D printing to fabricate the electrodes as a back end of line process.  This should allow all of the integration capabilities currently being developed by the community to be retained unaltered.

\subsection{Dielectric Material}

We propose using ormocer (\textbf{or}ganically \textbf{mo}dified \textbf{cer}amic), a silica-based material, as our printable dielectric material.  The high RF frequencies ($\sim$10-100\,MHz) and voltages ($\sim$10-1000\,V) applied to ion trap electrodes places stringent requirements on the electrical and thermal properties of any dielectrics used.  High dielectric strength is required to avoid breakdown.  High resistivity and low loss tangent avoid direct heating of the dielectric.  A low dielectric constant reduces parasitic capacitance, requiring less current to drive the electrodes, and thus lower Ohmic losses in the electrode metal.   Finally, a high thermal conductivity ensures any heating from losses is minimized.  Ormocer properties are somewhat formulation dependent, but in Table~\ref{tab:ormocer-characteristics} we give some representative values measured in the literature to give an idea of what can be achieved.

% along with the same values for SiO$_2$, a common ion trap dielectric, for comparison,

\begin{table}
\centering
\begin{tabular}{|c|c|} 
\hline
Relative permittivity & 2.5~\cite{Froehlich02}, 2.7~\cite{Laine92}, 3.1~\cite{Johansson03}  \\
\hline
Loss tangent & 0.0035~\cite{Johansson03}, 0.00397~\cite{Laine92}, 0.004~\cite{Froehlich02},\\
\hline
Dielectric strength (V/cm) & $8.7\times10^5$~\cite{Laine92}  \\
\hline
Bulk resistivity ($\Omega\cdot\text{cm}$) & $4.5\times 10^{16}$~\cite{Laine92}  \\
\hline
Thermal conductivity (W/m$\cdot$K) & 2.3\cite{Laine92}   \\
\hline
\end{tabular}
\newline
\caption{Electrical and thermal properties of ormocer.}
\label{tab:ormocer-characteristics}
\end{table}

For room temperature operation, outgassing needs to be low to ensure ultra-high vacuum pressures can be reached and the material must be able to withstand $\sim$200\,$^\circ$C vacuum bakes.  Low outgassing is also desirable as it reduces the chance of contaminating metal electrode surfaces, which are known to play a role in anomalous heating~\cite{hite2013surface}.  The organic cross-linked structure of ormocer reduces outgassing and makes ormocer fit for vacuum processes~\cite{fraunhofer}, although quantitative outgassing data is so far lacking.  Ormocer is stable to $>300\,^{\circ}$C~\cite{fraunhofer}.  For cryogenic operation, vacuum compatibility is much less of an issue but the structure must be able to survive repeated temperature cycles without structural failure, something that will need testing in future work.

\subsection{Initial Fabrication}

\subsubsection{Print}
\begin{figure}[h]
\includegraphics[width=0.95\textwidth]{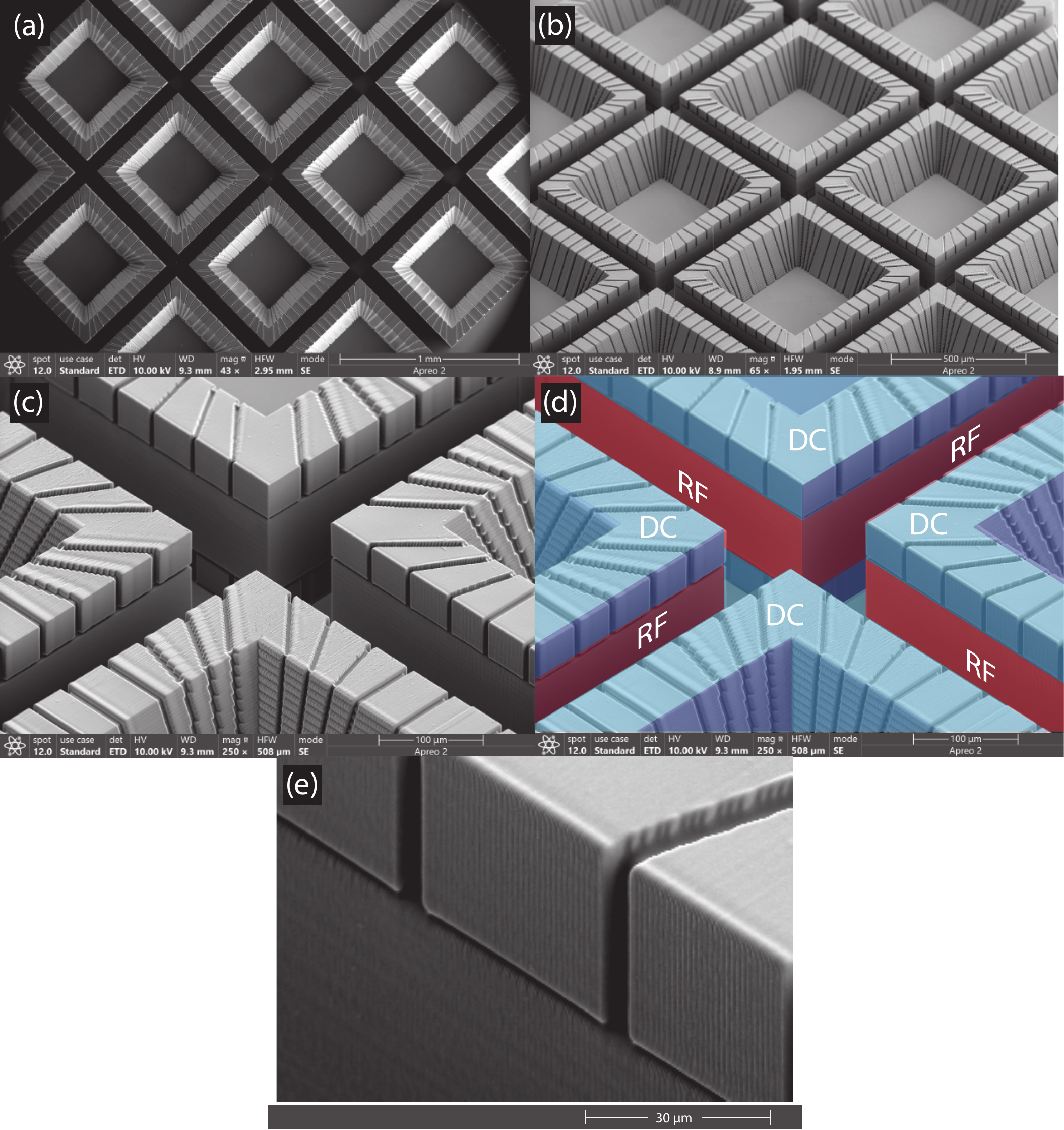}
\centering
\caption{\label{unmetalized-prints}SEM images of gold coated ormocer trap structures printed on glass. In this structure, many trenches have been connected up with X-junctions, as is envisaged in a large `QCCD' processor~\cite{Blakestad09}. (a) An overhead view of a grid of symmetric stacked trenches showing segmentation of the top DC electrodes.  (b) An oblique view of the same grid.  (c) A close up view of an intersection of trenches.  (d) A false-color version of (c) showing DC electrodes in blue and RF electrodes in red.  (e) A close-up view of a DC electrode segment.}
\end{figure}

As an initial test, ormocer based trap structures were printed onto a borosilicate glass wafer and metalized with $\sim20$\,nm of sputtered gold.  SEM images of which are shown in Fig.~\ref{unmetalized-prints}.  The printed structures are  $200\,\mu$m high with $100\,\mu$m wide trenches in between them, giving a $50\,\mu$m ion-electrode distance.  This is a fairly typical value, with most microfabricated traps historically lying in the few-10s to few-100s\,$\mu$m range. These traps were fabricated via a two-photon direct laser write process. A custom printer was used, based on open source concepts from the Janelia's open source Modular In vivo Multiphoton Microscopy System (MIMMS) and the printer described in~\cite{Pearre2019}.  

This printer was built to address specific needs in terms of print speed and full wafer scalability while retaining the necessary resolution. In the print process, an individual 2D field of view (FOV) is scanned using a galvo for $y$ axis steps paired with a fast resonant imaging mirror operating at 8\,kHz in the $x$ axis; allowing for printing at speeds up to $\sim$8000\,mm/s. A piezoelectric actuator is then used to sweep this FOV in the normal $z$ axis to form an individual metavoxel. Series of these are then stitched together across the entirety of the wafer via a controllable hexapod (Physik Instrumente) with sub-micron precision. We chose to break each “island” into 4 metavoxels with a $200\,\mu$m~$\times$~$200\,\mu$m FOV. Each individual island is printed in under a minute with a voxel resolution under $2\,\mu$m in the $xy$ print plane (142  $x$ and  512 $y$ voxels) and under $3\,\mu$m in $z$. Demonstrating that, whilst this is an inherently serial  process, the overall  processing  times at the required resolution are  tractable  for  much  larger traps than are currently in use. 

%Fluorescein (Sigma) for in situ imaging during printing (Hearnisch et al 2015).

A hybrid resist is used in this printing process, based on Ormocomp photoresist (Microchem). To this we add a photoinitator, 2.4.6-trimethylbenzoyl phosphine oxide (TPO) (Sigma), a stabilizing agent, 3,5-Di-tert-butyl-4-hydroxytoluene (BHT), and Fluorescein (Sigma) for in situ imaging during printing. A Chameleon Discovery laser, set to 780\,nm, 100\,fs pulse width, 80\,MHz Rep rate, and $\sim$100\,mW was used to initiate polymerization. Resolution and therefore surface roughness is a function of resin choice as well as the fundamental optics and scan mechanics of the printer. The current design takes advantage of the critical electrical dimensions of the trap being on a vertical surface which minimizes discretization concerns. In addition, a small amount of shrinkage during curing is inevitable. We minimize this by using the highest possible laser power before explosions occur. We do not see delamination from the substrate or noticeable distortion of the structures under SEM imaging.

%As an initial test, ormocer trap structures were printed onto a glass  wafer and metalized with $\sim20$\,nm of sputtered gold.  SEM images of which are shown in Fig.~\ref{unmetalized-prints}.  The printed structures are $200\,\mu$m high with $100\,\mu$m wide trenches in between them (ie. a $50\,\mu$m ion-electrode distance).    

%Morgan to add details of:
%- Our specific ormocer formulation/preparation
%- Laser, wavelength, pulse length, energy **cite**
%- Spot size, resolution, and FOV used and the range this printer is capable of.
%- Comment on FOV stitching necesity and capabilities
%- Curing schedule
%
%Surface roughness - see fig. 8 bottom right.
%
%Each electrode 'island' took $\sim$1\,minute to print, demonstrating that, whilst this is an inherently serial process, processing times are tractable even for much larger traps than are currently in use. Print time scales linearly with the volume of the print. **what are trade-offs on resolution**
%
%A small amount of shrinkage during curing is inevitable.  We minimize this by using the highest possible laser power before explosions occur.  We do not see delamination from the substrate or noticeable distortion of the structures under SEM imaging.

\begin{figure}[h]
\includegraphics[width=0.95\textwidth]{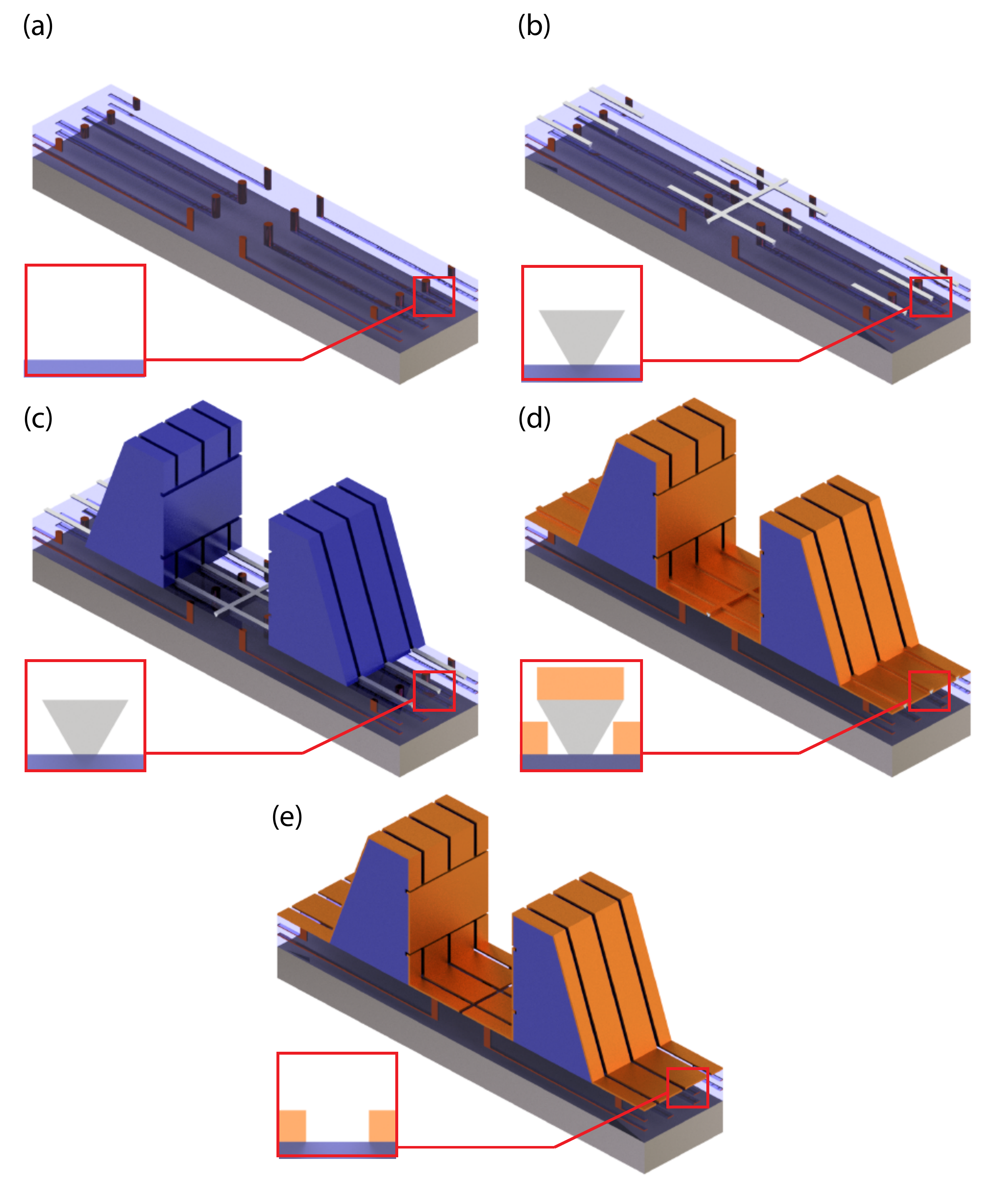}
\centering
%\caption{\label{fab-process}Proposed integration of electrode metalization with printing process shown here on a short section of the larger trap shown in Fig~\ref{unmetalized-prints}.  (a) Silicon wafer with electrode routing and vias up to the surface is fabricated using standard CMOS process (SiO$_2$ dielectric shown transparent).  (b) Aluminum is deposited and patterned to define gaps between electrodes.  (c) DLW used to print ormocer trap structures.  (d) Gold is deposition from multiple angles to coat all surfaces of the print.  (e) Aluminium is etched to allow liftoff of the gold over the gaps between electrodes.  Insets shown the liftoff process in more detail.}
\caption{\label{fab-process}Proposed integration of electrode metalization with printing process shown here on a short section of the larger trap shown in Fig~\ref{unmetalized-prints}.  (a) Silicon wafer with electrode routing and vias up to the surface is fabricated using standard CMOS process (metal shown copper coloured,  SiO$_2$ dielectric shown transparent).  (b) Aluminum is deposited and patterned to define gaps between electrodes.  (c) DLW used to print ormocer trap structures (blue).  (d) Gold is deposition from multiple angles to coat all surfaces of the print.  (e) Aluminium is etched to allow liftoff of the gold over the gaps between electrodes.  Insets shown the liftoff process in more detail.}
\end{figure}

\subsubsection{Metalization}

The next stage in our fabrication development is metalization.  Our trap structure is designed to be `self masking' - the isolation gaps between electrodes can be made deep enough (or with an undercut) such that directional deposition of metal onto the structure will not cause shorting between them. This metalization will connect the trap to vias already pre-fabricated on a planarized wafer using a standard CMOS process (see fig.~\ref{wafer}).  These vias must be off to the side of the printed structures as printing directly onto metal is difficult - the laser heats the metal causing violent disruption of the print.  On the substrate around the prints, isolation between electrodes will be achieved using a liftoff process.  We envisage using gold as our electrode metal as this is well established in ion trapping, and aluminium for the lifoff material.  The full process can be seen in Figure\,\ref{fab-process} and will be described in more detail in a forthcoming publication.

\subsubsection{Integrated Optics}

The final step to producing a working trap will be to integrate optics for laser delivery.  This is because trench designs do not offer the required optical access for free-space laser delivery.   Waveguides for light routing and grating couplers to focus beams onto the ions have now been demonstrated by several groups~\cite{Mehta2016, Ivory21, Mehta20}.  These demonstrations were using SETs, but one of the strengths of the traps proposed here is that the same technique can be used without modification. Other integrated features shown in Fig.~\ref{wafer}, whilst desirable, aren't an immediate requirement.

\section{Conclusion}

Additive 3D printing has the potential be a powerful enabling technology for trapped-ion quantum computing. It allows the combination of wafer-integrated ion control with the 3D structures required to obtain optimal trap performance. In order to realise this we have developed a novel trench geometry and shown it can be printed.  We lay out a roadmap to integrate the necessary electrical and optical routing to enable functional devices.

\section*{Acknowledgements}

We would like to acknowledge Aaron Casserly who suggested this collaboration, David Miller who assisted with process development, and Daniel Slichter who provided simulation code examples.\\

%#NSF for funding\\

%\clearpage

\FloatBarrier

\bibliography{efabrefs}
\bibliographystyle{unsrt}

\pagebreak

\end{document}